\documentclass{physeauth}
\usepackage{graphicx}
\usepackage{amsmath}
\usepackage{amssymb}

\begin{document}

\begin{frontmatter}

\title{Variational approach to the Caldeira-Leggett model}

\author[mpi,amst]{Javier Sabio \thanksref{thank1}} and
\author[amst]{Fernando Sols}

\address[mpi]{Instituto de Ciencia de Materiales de Madrid, Sor Juana In\'es de la Cruz 3, E-28049 Madrid, Spain}
\address[amst]{Departamento de F{\'\i}sica de Materiales, Universidad
  Complutense de Madrid, E-28040 Madrid, Spain.}

\thanks[thank1]{
Corresponding author.
E-mail: javier.sabio@icmm.csic.es}

\begin{abstract}
  We apply the displaced-oscillator variational ansatz to the Caldeira-Leggett model for a quantum particle in a one-dimensional box described by a tight-binding chain. We focus on the case of an Ohmic environment and study the phase diagram for different chain lengths. At zero temperature there is a phase transition to a localized phase when the number of sites is even. At finite temperature, a transition from a coherent to an incoherent regime is predicted for all the chain lengths considered. Finally, the results are compared to those obtained with numerical techniques.
\end{abstract}

\end{frontmatter}


\section{Introduction}

Open quantum systems have been widely explored in the literature. From the very beginning of Quantum Mechanics it was realized that every real quantum system is, to some extent, coupled to an environment. Hence, the investigation of the way this coupling can affect the properties of the system is crucial to understand the experimental information. Moreover, the study of open quantum systems has proved very useful to analyze the quantum origin of dissipation \cite{Weiss}. The concept of environment has become central in the understanding of how the classical world emerges from the quantum world \cite{Zurek}.

In order to analyze open quantum systems, simple models of bath and system-bath coupling have been proposed. A bath composed of an infinite sum of harmonic oscillators is known to provide a sufficient description to understand a wide spectrum of properties of those systems \cite{Feynman}. As Caldeira and Leggett realized in their seminal paper on quantum dissipation \cite{CL1}, to reproduce the classical equations of motion of a particle with friction it is enough to consider a linear coupling to its position. Since then, the Caldeira-Leggett model has become a fruitful tool to understand basic properties of the influence of the environment on a quantum particle in both dynamical and equilibrium contexts.

In particular, the case of a one-dimensional quantum particle coupled to an environment exhibits a rich physical behavior. Three simple situations can be considered. Firstly, that of a free particle coupled to the bath. In Ref. \cite{HakimAmbegaokar}, it was shown that the effect of the bath is an effective renormalization of the mass of the particle, that can be understood physically as the particle dragging a cloud of bath oscillators. This is actually equivalent to the interpretation found in the polaron problem \cite{Mahan}, where a related -but different- model is considered. The second case of interest is the particle in a periodic potential, studied in \cite{Schmid,Guinea1}. A remarkable difference with the first case is the existence of a quantum phase transition at zero temperature, in which the particle becomes localized. Finally, a particle in a confined region of space can be considered (the so-called quantum particle in a box). The lack of symmetries of this model makes it difficult to apply most of the standard techniques available for other problems. Recently, a discretized version of the particle-in-a-box problem has been studied with the numerical renormalization group (NRG), performing an analysis of the zero-temperature phase diagram \cite{Sabio}. It was found that the coupling to the environment gives rise to a narrowing of the spatial density distribution of the particle, until a critical value of the coupling is reached where a localization phase transition occurs.

In this paper we apply a variational ansatz to study the zero and finite temperature phase diagram of the dissipative quantum particle confined to a finite chain. The variational ansatz has been extensively used to analyze the celebrated spin-boson model (see \cite{RMP_spin_boson} for a general description), which actually turns out to be the two-site limit of the model considered here.  Here we will analyze the case of $M$ sites, complementing and clarifying previous results at zero temperature, already reported in \cite{Sabio,Erratum}, and adding the finite temperature calculation.

The paper is organized as follows: Section 2 describes the model, pointing out previous results obtained in the study of its equilibrium phase diagram at zero temperature. Section 3 contains the main results of the paper: the variational approach is introduced, and results are obtained for zero and finite temperature. Finally, section 4 contains a general discussion and a comparison with the NRG results.

\section{The model}
\label{sec:RW}

We analyze a finite tight-binding chain coupled to a dissipative bath in the way Caldeira and Legget envisaged:
\begin{eqnarray}
\mathcal{H} = \mathcal{H}_{\rm{kin}} + \mathcal{H}_{\rm{bath}} + \mathcal{H}_{\rm{int}} + \mathcal{H}_{\rm{ct}} \nonumber \\
\mathcal{H}_{\rm{kin}} = -t \sum_m^{M} (|m\rangle \langle m+1| + |m + 1\rangle \langle m|) \nonumber \\
\mathcal{H}_{\rm{bath}} = \sum_{k<\omega_c} k b_k^{\dagger} b_k \nonumber \\
\mathcal{H}_{\rm{int}} = \lambda q \sum_{k<\omega_c} \sqrt{k} (b_k^{\dagger} + b_k) \nonumber \\
\mathcal{H}_{\rm{ct}} = \lambda^2 q^2 \sum_{k<\omega_c} 1
\end{eqnarray}
The first term is the kinetic term of the tight-binding chain, $m$
labelling the different $M$ sites, and $t$ being the hopping between
sites. By taking $t = \hbar/2ma^2$, with $m$ the mass of the particle
and $a$ the space between sites, in the limit $a\rightarrow 0$,
$M\rightarrow \infty$ and $L = Ma$ constant, we recover the continuum
model of a confined quantum particle. The second term is that of a
gapless bath of harmonic oscillators. The distribution of momentum of
the oscillators can be taken continuous below a certain cutoff
$\omega_c$. The third term is the linear coupling of the bath to the
position $q = \sum_m^M (m-m_0) |m\rangle \langle m|$ of the particle,
$m_0$ being a label for the center of the chain. The coupling is
characterized by the bath spectral function $J(\omega) = \pi \lambda^2
\sum_{k<\omega_c} k \delta(\omega-k)$. In our particular case, we will
limit the discussion to an Ohmic bath, which fulfills the constraint
$J(\omega) = 2 \pi \alpha |\omega|\theta(\omega_c- \omega)$, with
$\alpha = \lambda^2/4\pi$. Finally, the last term is a counter-term
required to ensure a homogeneous dissipation along the chain despite
the coupling to the position, which is sufficient in equilibrium
contexts \cite{SanchezSols}. Actually, that the coupling does not
privilege any site can be seen by performing a unitary transformation
on the Hamiltonian, $U = e^{-\lambda q \sum_k (b_k^{\dagger} - b_k)/\sqrt{k}}$. As a result,
\begin{equation}
\mathcal{\tilde{H}} = \sum_{k<\omega_c} k b_k^{\dagger} b_k - t\sum_m^M[|m\rangle \langle m+1| e^{-\lambda \sum_k (b_k^{\dagger}-b_k)/\sqrt{k}}+\rm{H.c.}] \label{H2}
\end{equation}
Notice that this transformation diagonalizes exactly the Hamiltonian when $t=0$.
An important remark concerning this model is that it only preserves parity symmetry.

As mentioned in the introduction, this model has been studied in Ref. \cite{Sabio} by using a version of the NRG technique conveniently suited to deal with bosonic baths \cite{Bulla}. Two remarkably results were found. (i) For small hopping (as compared to the cutoff, $t\ll1$), and couplings smaller than a certain critical coupling of order $\alpha \sim 1$, the coupling to the bath increasingly confines the particle to the center of the chain; i.e., despite of the homogeneous coupling, the boundary conditions are responsible for a non-homogenenous effect. (ii) For couplings larger than the critical value a new phase of the model was found, in which the particle gets localized at one single site, breaking the parity symmetry. However, not every site is permitted, only two sites localized symmetrically with respect to the center but never at the edges. In a naive analysis of the model an equal probability of getting localized at every site of the chain would have been expected. Thus, there is a second non-homogeneous effect coming from the presence of boundaries that shows up in the localized phase.

\section{Variational approach}

We will use the approach proposed by Silbey and Harris \cite{Silbey} to study the spin-boson model, and later extended and discussed in a variety of papers \cite{Chen,Legg91,Chin}. The idea is to generalize the unitary transformation used in the previous section to get Hamiltonian (\ref{H2}) by the introduction of a variational parameter $f_k$:
\begin{equation}
U = e^{q \sum_k (f_k/k) (b_k^{\dagger} - b_k)}
\end{equation}
As this kind of transformation displaces the oscillators from their unperturbed equilibrium positions, such a variational approach is conventionally referred to as the displaced oscillator ansatz. The transformed Hamiltonian reads:
\begin{eqnarray}
\mathcal{\tilde{H}} = U \mathcal{\tilde{H}} U^{\dagger} = \mathcal{\tilde{H}}_0 + \mathcal{\tilde{H}}_{\rm{res}} \nonumber \\
\mathcal{\tilde{H}}_0 = -t_{\rm{ren}} \sum_m^M (|m\rangle \langle m+1| + H.c) + \sum_{k<\omega_c} k b_k^{\dagger} b_k + g q^2 \nonumber \\
\mathcal{\tilde{H}}_{\rm{res}} =  V_{+} \sum_m^M |m+1\rangle \langle m| + V_{-} \sum_m^M |m\rangle \langle m+1| + V_0 q \nonumber \\
\end{eqnarray}
where we have defined the potential strength $g = \sum_k (\lambda + \frac{f_k}{\sqrt{k}})^2$ and a renormalized hopping given by $t_{\rm{ren}} = \langle te^{-\sum_k (f_k/k)(b_k^{\dagger}-b_k)}\rangle_T$, with the thermal average taken with respect to $\mathcal{\tilde{H}}_0$. The latter can be worked out analytically, yielding $t_{\rm{ren}} = t \exp[-\frac{1}{2} \sum_k \frac{f_k^2}{k^2} \coth(\frac{\beta k}{2})]$. Such a separation of the Hamiltonian ensures that the operators contained in the residual term $\mathcal{\tilde{H}}_{\rm{res}}$,
\begin{eqnarray}
V_{+} = V_{-}^{\dagger} = t e^{-\sum_k \frac{f_k}{k}(b_k^{\dagger} - b_k)} - t_{\rm{ren}} \\
V_0 = \sum_{k<\omega_c} (\lambda \sqrt{k} + f_k)(b_k^\dagger + b_k) \, ,
\end{eqnarray}
have zero thermal expectation values, $\langle V_i \rangle_T = 0$  ($i=\pm,0$). Following Silbey and Harris \cite{Silbey}, we compute the Bogoliubov-Feynman upper bound of the free energy \cite{FeynmanBook}:
\begin{equation}
A_B = -\beta^{-1} \log \rm{Tr} ( e^{-\beta \mathcal{\tilde{H}}_0}) + \langle \mathcal{\tilde{H}}_{\rm{res}}\rangle_T + \mathcal{O}(\langle \mathcal{\tilde{H}}_{\rm{res}}^2\rangle_T)
\end{equation}
A first relevant difference with the spin-boson model, is that now the Hamiltonian $\mathcal{\tilde{H}}_0$ with which the thermal averages are taken, contains a potential term quadratic in the position of the particle, with potential strength $g>0$. The effect of such a term is a certain localization of the particle at the center of the chain, making the Hamiltonian not easy to diagonalize, as the actual eigenenergies and eigenstates depends on $t_{\rm{ren}}$ and $g$. While a closed expression for $A_B$ cannot be worked out,  the condition for a minimum does give rise to the following self-consistent equation for the variational parameter $f_k$:
\begin{equation}
f_k = \frac{-\lambda k^{3/2}}{k- \frac{1}{2}\frac{\langle\mathcal{\tilde{H}}_{0,\rm{kin}}\rangle_T}{\langle q^2 \rangle_T} \coth(\frac{\beta k}{2})}
\end{equation}
where $\mathcal{\tilde{H}}_{0,\rm{kin}} \equiv -t_{\rm{ren}} \sum_m (|m\rangle \langle m+1| + \rm{H.c.})$. Plugging this expression into the definitions of the renormalized hopping and the potential strength, we can get self-consistent equations for the actual parameters of the effective Hamiltonian. If we define $s_T(t_{\rm{ren}}, g) \equiv \left|\frac{1}{2}\frac{\langle\mathcal{\tilde{H}}_{0, kin}\rangle_T}{\langle q^2 \rangle_T}\right|$, we obtain
\begin{eqnarray}
t_{\rm{ren}} \equiv t e^{-B(t_{\rm{ren}})} \nonumber \\
B(t_{\rm{ren}}) = \alpha \int_0^{\omega_c} d\omega \frac{\omega \coth(\frac{\beta \omega}{2})}{[\omega - s_T \coth(\frac{\beta \omega}{2})]^2}  \\
g(t_{\rm{ren}}) = \frac{\alpha}{2}  \int_0^{\omega_c} d\omega \left[ \frac{s_T \coth(\frac{\beta \omega}{2})}{\omega - s_T \coth(\frac{\beta \omega}{2})}\right]^2
\end{eqnarray}
The study of the variational solution requires the numerical solution of these two equations. Starting from a seed, Hamiltonian $\mathcal{\tilde{H}}_0$ is diagonalized to calculate the thermal expectation values and the corresponding renormalized hopping and potential strength.

As pointed out in previous works, the variational solution regulates the strength of the coupling between bath modes and the particle in a way that preserves the spirit of the adiabatic approach \cite{RMP_spin_boson}, but at the same time it takes control of the infrared divergences introduced by the slow modes of the bath. This sort of separation between fast and slow modes is one of the main virtues of the approach.

\subsection{Zero temperature}

A further insight into the solution can be obtained at zero temperature, i.e., when $\beta \rightarrow \infty$. Then the equations for the Hamiltonian can be integrated analytically, yielding:
\begin{eqnarray}
t_{\rm{ren}} = t\left(1 + \frac{1}{s_0}\right)^{-\alpha} e^{\frac{\alpha}{s_0+1}} \label{sce2}\\
g = \frac{2 \alpha s_0}{s_0 + 1}
\end{eqnarray}
where we have taken $\omega_c = 1$ for simplicity. The free energy bound reduces to an energy bound, which can also be worked out analytically:
\begin{equation}
A_B(T = 0) = E_B = \langle\mathcal{\tilde{H}}_{0,{\rm kin}}\rangle_0 + g\langle q^2 \rangle_0
\end{equation}

For every $\alpha$ and $t$ we have a trivial solution of the equations, corresponding to $t_{\rm{ren}} = 0$ and, by virtue of their relation, $g = 0$. This would correspond to a particle localized at any site of the chain with a cloud of oscillators dressing it. An interesting implication comes from the relation between $g$ and $t_{\rm{ren}}$. As $g$ controls the non-homogeneous effects induced by the environment in the density profile, no localized solution coming from the variational approach will show preference for a particular site in the chain, as we mentioned it happened in the numerical solution. We will discuss this later.

\begin{figure}[tb]
  \begin{center}
    \includegraphics[angle=0,width=.47\textwidth]{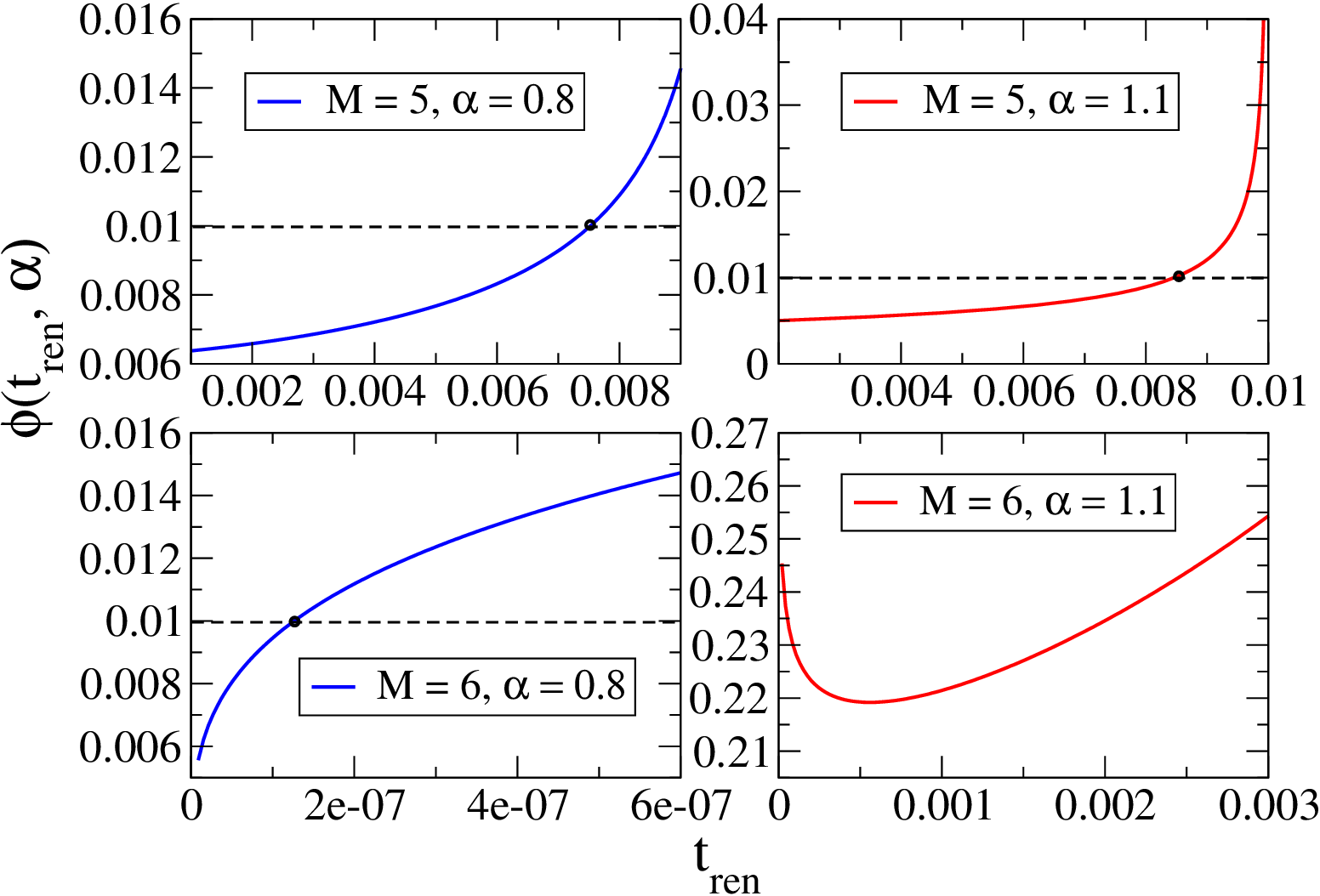}
    \caption{Graphical solution of the self-consistent equation (\ref{sce2}). The function plotted is defined as $\phi(t_{\rm{ren}}, \alpha) \equiv t_{\rm{ren}}(1 + 1/s_0)^\alpha e^{-\alpha/(s_0+1)}$, where $s_0 = s_0(t_{\rm{ren}})$. The condition for non-trivial solutions is thus $\phi(t_{\rm{ren}}, \alpha) = t$. In the figures we have chosen two representative cases of odd and even chains, $M = 5, 6$. The value of the hopping is $t = 0.01$.}
    \label{fig1}
  \end{center}
\end{figure}

\begin{figure}[tb]
  \begin{center}
    \includegraphics[angle=0,width=.47\textwidth]{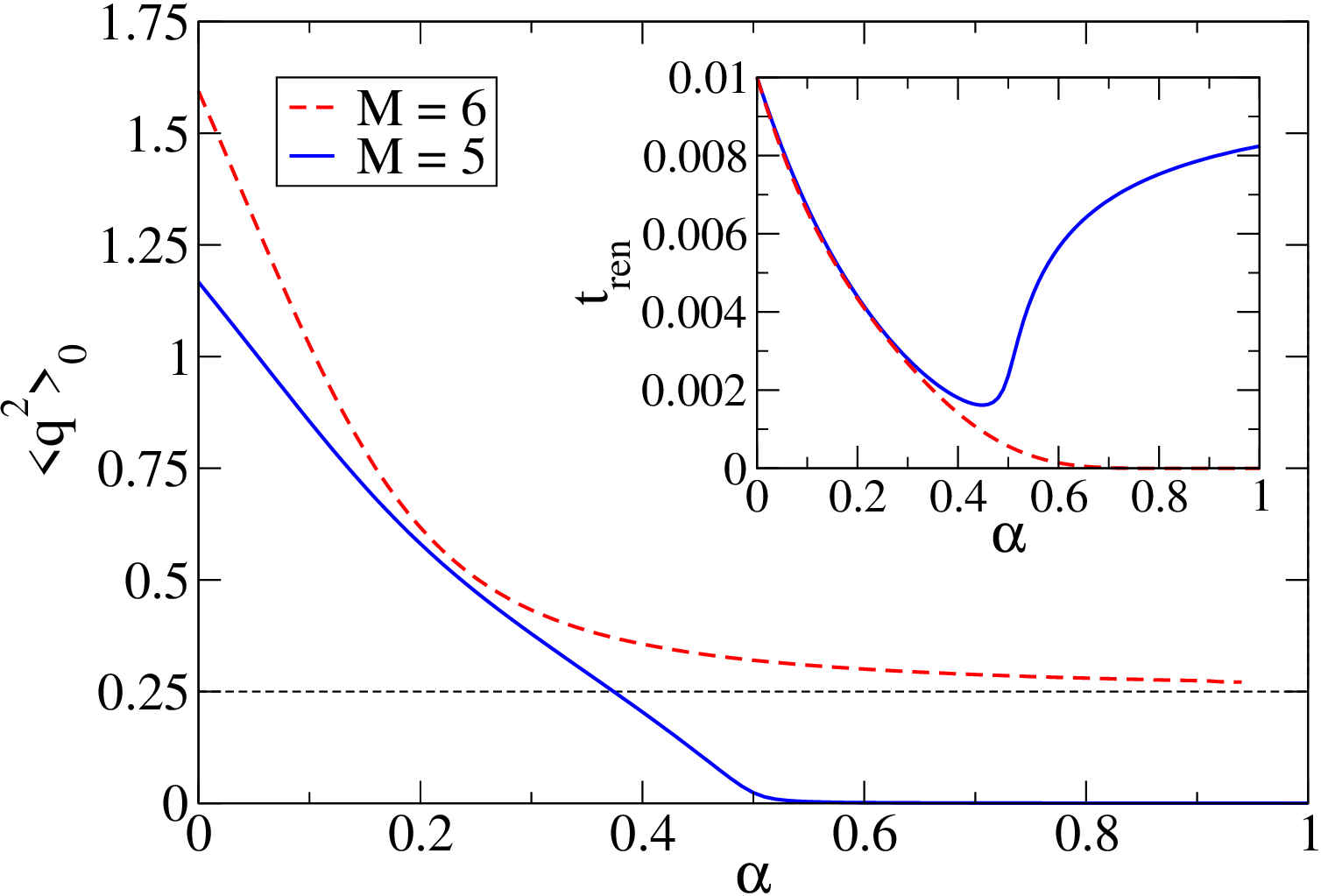}
    \caption{Mean squared position deviation for the dissipative confined particle in chains of $M = 5$ and $6$ sites. In both cases the coupling to the environment leads to a narrowing of the density distribution of the particle towards the center. Here, $\langle q^2 \rangle_0 = 0.25$ is the mean squared position corresponding to a two-site system. Notice that, for odd chains, there is a crossover to a region in which the particle is effectively localized at the central site. Inset: renormalization of the hopping, $t_{\rm{ren}}$, as a function of the coupling to the environment, also for $M = 5$ and $M = 6$. The crossover for odd chains is reflected in the behaviour of the renormalized hopping.}
    \label{fig2}
  \end{center}
\end{figure}

The energy associated to the trivial solution is zero. The existence of non-trivial solutions of the equations with lower energy determines the phase diagram at $T = 0$, in terms of the number of sites $M$ and the parameters $t$ and $\alpha$. As shown in Fig. \ref{fig1}, the region where there are non-trivial solutions with lower energy depends on the number of sites being even or odd. For odd chains there is a non-trivial solution with finite $t_{\rm{ren}}$ and $g$ for every value of $\alpha$. In this solution, the potential induced by the environment confines increasingly the particle at the center, having in fact a crossover to a sector in which the particle at all instances is localized at the center (Fig. \ref{fig2}). Such a crossover is also reflected in the renormalized hopping (inset of Fig. \ref{fig2}), as in this region the hopping gets less renormalized with increasing coupling to the bath. For even chains there is no longer a non-trivial solution for $\alpha > 1$, and a phase transition occurs to a localized phase in which the parity is broken. In the delocalized phase ($\alpha < 1$), the presence of the environment results also in a certain confinement of the particle to the two central sites in this case. The hopping is further renormalized as the coupling increases, an no crossover is observed. We will discuss this effect in the last section of the paper.

\subsection{Finite temperature}

\begin{figure}[tb]
  \begin{center}
    \includegraphics[angle=0,width=.47\textwidth]{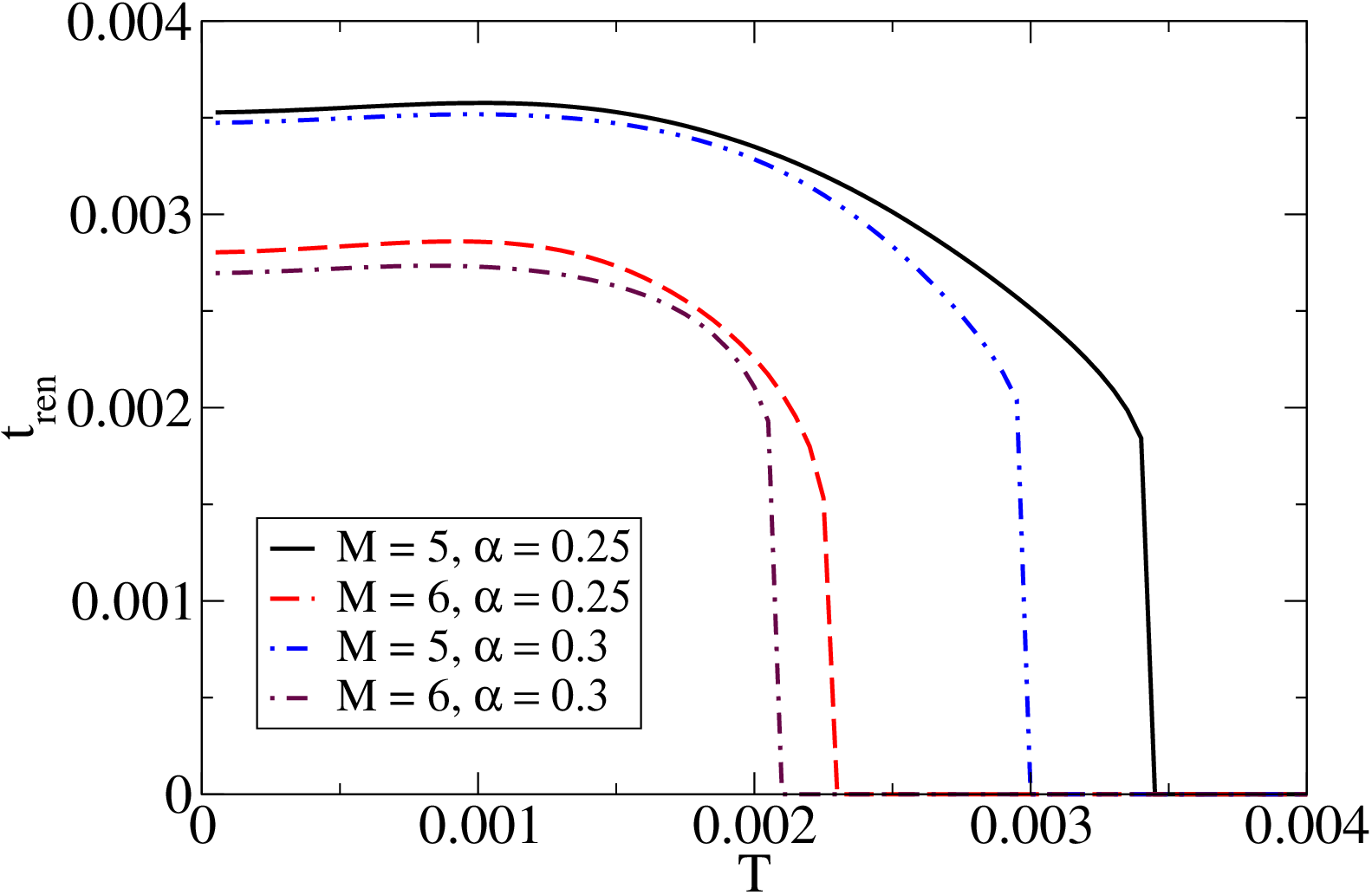}
    \caption{Dependence of the renormalized hopping $t_{\rm{ren}}$ on the temperature, for two representative cases of odd and even chains, $M = 5$ and $M = 6$, respectively. For $\alpha < 1$ in both cases there is a critical temperature that separates an incoherent high temperature regime, where $t_{\rm{ren}} = 0$, and a coherent low temperature one, where $t_{\rm{ren}}$ is finite. Notice that, for low temperatures, the renormalized value of the hopping grows with temperature.}
    \label{fig3}
  \end{center}
\end{figure}

For finite temperatures, in general, the self-consistent equations have to be integrated and solved numerically. Again, the results depend on the number of sites of the chain being even or odd. In both cases, for couplings $\alpha < 1$,  we find a phase transition at some critical temperature $T^*$ from a coherent regime where $t_{\rm{ren}}$ and $g$ are finite, to an incoherent high-temperature region where $t_{\rm{ren}} = g = 0$. In this phase, the Hamiltonian has degenerate energy levels and they are equally occupied in a statistical thermal mixture. In Fig. \ref{fig3} we see a typical plot of the behavior of $t_{\rm{ren}}(T)$ as a function of the temperature. Notice that for small coupling to the bath, $\alpha \ll 1$, and high temperatures, $T \gg t_{\rm{ren}}$, we have a trivial expression for the parameter 
\begin{equation}
s_0 = \left| \frac{1}{2} \frac{\langle\mathcal{\tilde{H}}_{0, kin}\rangle_0}{\langle q^2 \rangle_0}\right| = \frac{M \beta t^2_{\rm{ren}}}{\sum_m^M (m-m_0)^2}\,\,.
\end{equation} 
This is the actual dependence found in \cite{Chin} for the spin-boson model in the same regime, generalized trivially to the case of $M$ sites by adding the factor $M/\sum_m(m-m_0)^2$. As stated already in this reference, the critical temperature can be worked out analytically and has the form:
\begin{equation}
T^* \sim \frac{t_{\rm{ren}}}{\alpha}\left[\frac{M}{\sum_m^M (m-m_0)^2}\right]^{1/2}\,\, .
\end{equation}
Another interesting issue is that, at low temperatures, the renormalized hopping $t_{\rm{ren}}$ gets larger, instead of smaller (see Fig. \ref{fig3}). This result was also observed in the Ohmic and sub-Ohmic spin-boson model \cite{Weiss,Kehrein,Chin}, and it is associated to the fact that, at low temperatures, only slow bosonic modes are excited, which increases the hopping (as opposed to the effect of fast modes).

\begin{figure}[tb]
  \begin{center}
    \includegraphics[angle=0,width=.47\textwidth]{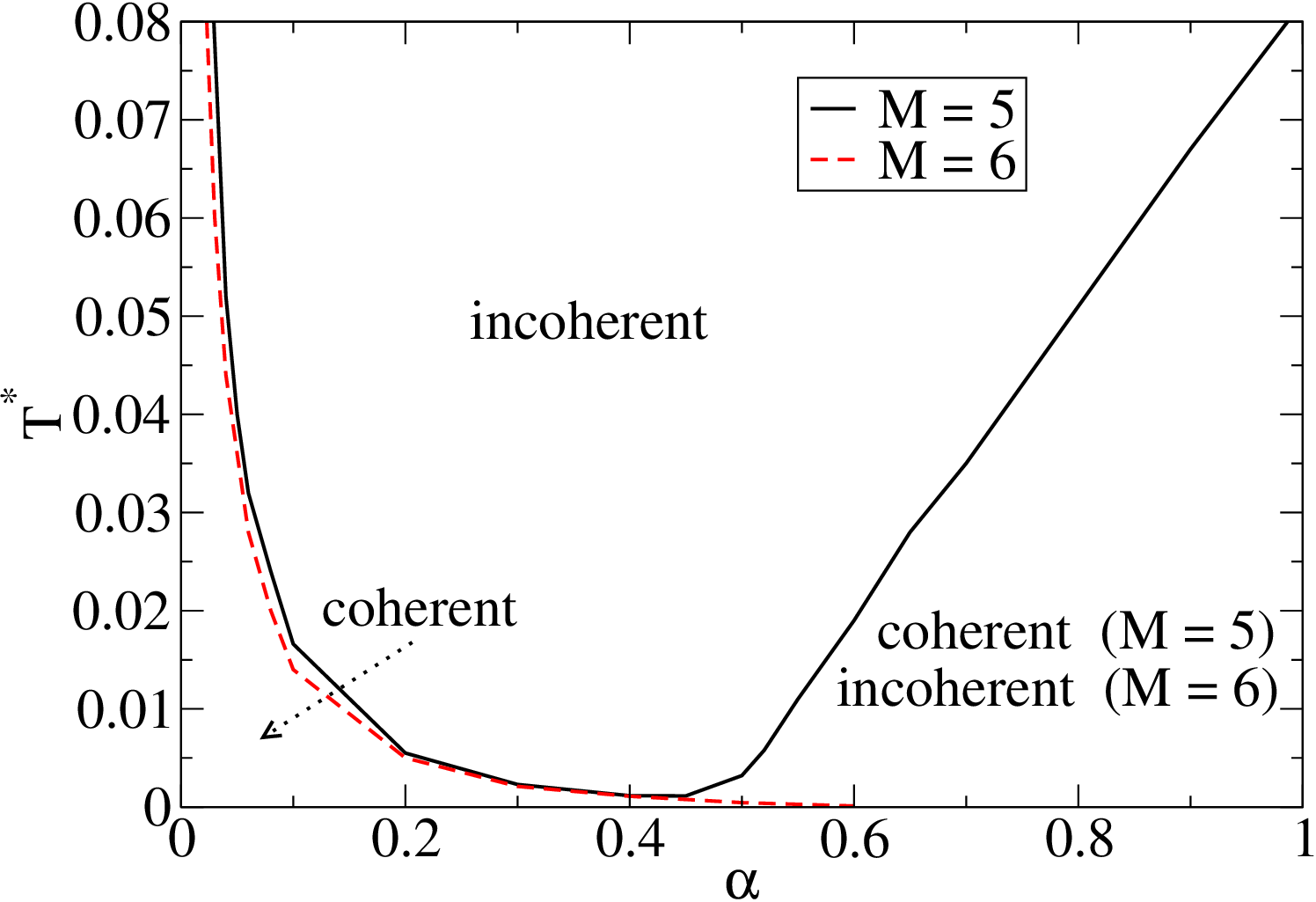}
    \caption{Finite temperature phase diagram of the model. The behaviour is different for odd and even chains. In both cases there are two phases, one where $t_{\rm{ren}} = 0$ (incoherent) and other where $t_{\rm{ren}}$ is finite (coherent). For small coupling $\alpha$ to the bath, the behaviour of the critical temperature is independent of the chain being odd or even. For large couplings the behaviour is dramatically different, and the coherent phase only prevails for $\alpha > 1$ for chains with an odd number of sites.}
    \label{fig4}
  \end{center}
\end{figure}

In the case of even chains, for $\alpha > 1$ there is no such a phase transition, and only the incoherent phase prevails. For an odd number of sites, the phase transition still occurs for arbitrary $\alpha$, a reminiscence of the lack of phase transition at $T=0$, that we discussed in the previous section. A phase diagram with these features is shown in Fig. \ref{fig4} for the two representative cases of $M = 5,6$.

\section{Discussion}

The variational calculation predicts a non-trivial phase diagram both at zero temperature and for finite temperatures. Being a sort of sophisticated mean-field calculation, it gives an effective Hamiltonian for a quantum particle with renormalized parameters $t_{\rm{ren}}$ and $g$ due to the coupling to the bath. Hence, two main effects arise from the environment: the renormalization of the hopping, and the confinement at the center of the array due to a quadratic potential. The results have a dependence on the number of sites in the array, with qualitative differences when odd or even chains are considered.

Let us understand those results. At zero temperature, there is no phase transition when the number of sites is odd. When the number of sites is even, the results are remarkably close to those found for the spin-boson model. This is not casual. As shown in Fig. \ref{fig2}, due to the quadratic potential, the particle is nearly almost localized at the two central sites for coupling strengths close to $\alpha = 1$. Thus, we have an effective spin-boson model, and the phase transition is essentially the same found in this model. Such an argument also applies to the finite temperature case,  where again we find results close to those predicted by the variational calculation in the spin-boson model.

In the case of an odd number of sites in the chain the situation is very different: here, at zero temperature, the quadratic potential confines the particle around the single central site for enough large couplings, and no effective spin-boson model physics arises. Once the particle is effectively located at the center, the renormalization of the hopping reverse its tendency, as shown in Fig. \ref{fig2}, increasing again. An explanation for this effect would come from the fact that the bath is not coupled to the central site, as here $\hat{q}|m_0\rangle = 0$. Thus the renormalization of the particle parameters would be smaller as the particle is more localized at the center. This behaviour is reflected in the finite temperature phase diagram: the coherent-incoherent transition being related to the renormalization of the coupling by the bath, the critical temperature starts to increase again. For very large coupling to the environment, it can be seen that the renormalized hopping tends to its bare value.

The results can be compared with the NRG results obtained in \cite{Sabio}. Both approaches qualitatively agree in the predictions for the mean squared position of the particle in the delocalized phase, in the sense that the particle becomes localized at the center, though the variational calculation shows a faster degree of localization as a function of the coupling strength than the numerical calculation \cite{Erratum}.

More relevant is the phase transition found, with NRG, at $\alpha \sim 1$, independently of the length of the chain, and where the particle becomes localized in preferred sites. A result that disagrees with the phase transition predicted in the variational calculation. A way to reconcile both approaches is to suppose that the effective Hamiltonian for the quantum particle is missing higher order contributions to the potential generated by the environment. Particularly, the simple potential $V(q) = g_1 q^2 + g_2 q^4$ could explain the numerical results just having $g_1 \propto \alpha_c - \alpha$ and $g_2 >0$. With this dependence, the potential would have a minimum at the center for $\alpha < \alpha_c$, and two minima symmetrically located around the latter when $\alpha > \alpha_c$. Then the nature of the phase transition would be more complicated than that found in the spin-boson model, as it would require the assistance of the potential generated by the coupling. Besides, this would predict a non-homogeneous density pattern in the localized phase, as contrary to what is found in the variational calculation. So far, however, more complicated generalizations of the variational ansatz have failed to provide such an effective potential, and further research is needed in order to clarify this point.

\section*{Acknowledgements}

The authors are grateful to F. Guinea for a critical reading of the manuscript, and A. J. Leggett for his useful insights. This work was supported by MEC (Spain) through
grants FIS2005-05478-C02-01, FIS2007-65723 and EU Marie Curie RTN Programme no. MRTN-CT-2003-504574. J.S. wants to
acknowledge the I3P Program from the CSIC for funding.


\end{document}